\documentclass[12pt]{iopart}
\usepackage{graphicx}
\usepackage[outdir=./]{epstopdf}
\usepackage{subfig}
\expandafter\let\csname equation*\endcsname\relax
\expandafter\let\csname endequation*\endcsname\relax
\usepackage{amsmath}
\usepackage{float}
\usepackage{xcolor}
\graphicspath{ {./figures/} }
 
\begin{document}

\title[Inter-qubit interaction in a linear array of three-dimensional cavities]{Inter-qubit interaction mediated by collective modes in a linear array of three-dimensional cavities}

\author{Dmytro Dubyna and Watson Kuo}

\address{Department of Physics, National Chung Hsing University, 250 Kuo-Kuang Rd., 402 Taichung, Taiwan}
\eads{\mailto{dubyna2005@ukr.net} and \mailto{wkuo@phys.nchu.edu.tw}}

\begin{abstract}
A design of LEGO-like construction set that allows assembling of different linear arrays of three-dimensional (3D) cavities and qubits for circuit quantum electrodynamics (cQED) experiments has been developed. A study of electromagnetic properties of qubit-3D cavity arrays has been done by using high frequency structure simulator (HFSS). A technique for estimation of inter-qubit coupling strength between qubits embedded in different cavities of cavity array, which combines Hamiltonian description of the system with simple HFSS simulations, has been proposed. A good agreement between inter-qubit coupling strengths, which were obtained by using this technique and directly from simulation, demonstrates the suitability of the method for more complex qubit-cavity arrays where usage of finite-element electromagnetic simulators is limited.
\end{abstract}

\vspace{2pc}

\section{Introduction}

Circuit quantum electrodynamics (cQED) studies light-matter interaction between an artificial atom (qubit) and a coplanar or three-dimensional (3D) waveguide cavity. Nowadays, cQED is widely used in quantum computation~\cite{RN1,RN2,RN3,RN16} and quantum simulation~\cite{RN4,RN5} where qubit and cavity serve as building blocks for creating complex qubit-cavity arrays. In most of these arrays, preference has been given to on-chip coplanar cavities that facilitate scalability of the sample. However, coplanar cavities have low mode volume and surrounded by different sources of energy dissipation due to the wiring, substrate, radiation etc. In addition, a dense location of elements on a chip leads to appearance of unwanted crosstalk~\cite{RN3}. All these factors hinder measurement and affect qubit performance. A good alternative to coplanar cavity is 3D cavity, which has much higher mode volume and makes the qubit better isolated from the environment. As a result, qubits in 3D cavity demonstrate a significant improvement in lifetimes~\cite{RN6,RN7}. In view of this, investigation of qubit-3D cavity arrays is of high scientific relevance.

One of the challenges in qubit-3D cavity arrays is to provide a good coupling between qubits embedded in different cavities of array. Ideally, many properties of the qubit-cavity arrays could be predicted even before sample fabrication by using finite-element electromagnetic simulator such as high frequency structure simulator (HFSS)~\cite{RN8} and  black box quantization technique (BBQ) proposed in~\cite{RN9}. However, with an increase of complexity of simulated model (e.g. introducing additional cavities and qubits in to array), the HFSS simulation becomes more time- and resource-consuming. This disadvantage can lead to the limits where HFSS simulation could be applied, forcing to search new solutions of the problem~\cite{RN10}.

It is known, that in the absence of direct coupling between distant qubits embedded in a single cavity, inter-qubit coupling could be achieved via exchange of a virtual photon with one of the cavity’s resonant frequency~\cite{RN11,RN12}. At the same time, when two or more identical cavities are coupled, their individual resonant frequencies are transforming in to the collective oscillations of the coupled system, which are called normal modes. In cavity array that consists of $N$ identical cavities, there would be $N$ normal modes in the vicinity of corresponding resonant frequency of the single cavity. Thus, for qubits located in different cavities of qubit-cavity array, inter-qubit coupling could be mediated by those normal modes. Inter-qubit coupling mediated by normal modes of 3 coupled coplanar resonators was studied in~\cite{RN13} where Hamiltonian describing the system was proposed. According to it, inter-qubit coupling depends on coupling strength between qubits and cavities to which they are directly coupled, detuning between qubits and cavity mode that mediates the coupling and on inter-cavity coupling strength. It turns out that these parameters could be found from a series of HFSS simulations involving simple models, which consist of one or two cavities and only one qubit (if necessary). Knowing these parameters, we can diagonalize Hamiltonian and estimate inter-qubit coupling strength.

In this paper, we have developed qubit-3D cavity arrays for realizing cQED experiments and studied their electromagnetic characteristics by using HFSS. Based on Hamiltonian proposed in~\cite{RN13} and simulated data, inter-qubit coupling for qubits located in different cavities of 3 coupled 3D cavities was found. These values were compared with inter-qubit coupling, which was obtained directly from HFSS simulations of the qubit-3D cavity array. A good agreement between both results demonstrates a perspective to use this approach for estimation of the inter-qubit coupling in more complex qubit-cavity arrays where usage of finite-element electromagnetic simulators is limited.

\section{Methods}

\begin{figure}
\includegraphics[width=\linewidth]{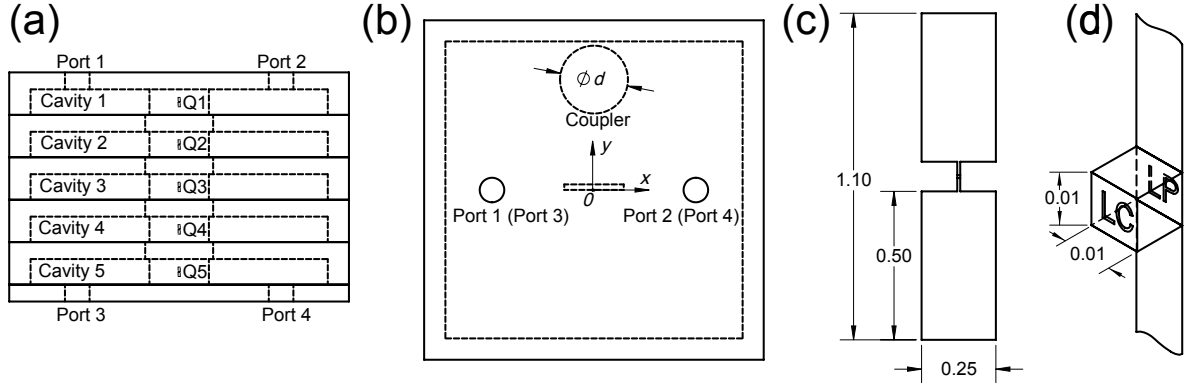}
\caption{
(a) Side and (b) top view of the cavity array assembly. Individual plates are forming five cavities that are connected through a hole (coupler). In each cavity, substrate with a qubit (Q1–Q5) is installed. (c) 3D model of the qubit that was used for simulation. (d) Isometric view of the feeding point of the dipole antenna. In order to obtain impedance (admittance) data of simulated system, a square patch at the antenna feeding point was assigned as a lumped port (LP).  A linear part of Josephson junction (JJ) was simulated by assigning a square patch, connected in parallel to the antenna feeding point, as an LC circuit (LC). All dimensions are in mm. 
}\label{fig:fig1}
\end{figure}

Figures~\ref{fig:fig1}a,~\ref{fig:fig1}b show a possible assembly of linear 3D cavity array for cQED experiments. The structure consists of five coupled cavities (Cavity 1--5) and five chips with qubits (Q1--Q5) inside of each cavity of the array. The base element of the assembly is metallic plate with a square pit. The cavities are formed by stacking the plates one by one. The inner plates of the array have a hole (coupler), which provides coupling with a neighboring cavity. Each terminated plate of the array has two holes where SMA ports (Port 1--4) for generation and readout microwave signals could be installed. The structure allows an easy way to assemble many different combinations of linear qubit-cavity arrays, like in a LEGO construction set. This feature is very useful for quantum simulations of one-dimensional lattices~\cite{RN5} where adding or subtracting elements in the chain could be done without affecting properties of the rest of the structure. 

Electromagnetic properties of qubit-3D cavity arrays were studied by using HFSS software. For this purpose, different 3D models of either single cavity or cavity arrays were built. In all models, cavities have the shape of a cuboid with width and length of 35 mm~$\times$~35 mm. These dimensions were chosen in order to obtain resonant frequency of transverse electric fundamental mode for individual cavity (TE101) equal to 6~GHz. Cavity arrays with cuboid height either $h1$=1.5 mm or $h2$=3 mm were studied. The cavity array with $h1$ was used only for obtaining data presented in Figure~\ref{fig:fig2}. For all other cases, cavities with $h2$ were simulated. Small cuboid heights were chosen in order not to take much space when plates are stacking up and high enough for installation of a real substrate into the cavity. The thickness of the inner wall between adjacent cavities was equal either to 1.5 mm or 2 mm for $h1$ or $h2$ cavity heights, respectively. The diameter of the holes for measurement ports was equal to 2.9 mm and their center-to-center distance was equal to 24 mm.  Space inside the structure was assigned as a vacuum while all the walls were assigned as a perfect conductor. The holes on the surface of terminated cavities were assigned as wave ports according to notations in Figure~\ref{fig:fig1}a,~\ref{fig:fig1}b. The appropriate coupler hole diameter ($d$) and position were subjects of study and they were chosen depending on simulation. The coupler position is described by $(x, y)$ coordinates in the units of (mm, mm) according to the coordinate system in Figure~\ref{fig:fig1}b. The coordinates were changed only in the first quadrant of the cavity array due to the symmetry of the structure, i.e. $(0, 0)$ coordinates correspond to the position of the coupler at the cavity center.

In order to simulate qubit behavior, a 3D model, which is shown in Figures~\ref{fig:fig1}c,~\ref{fig:fig1}d, was placed inside the cavity. The structure was built by using patch objects and consists of two paddles of dipole antenna, which are well seen in Figure~\ref{fig:fig1}c. The pads dimensions are typical for transmon qubit~\cite{RN15} in 3D cavity experiments~\cite{RN6}. For obtaining impedance (admittance) data of simulated system, a square patch at the antenna feeding point was assigned as a lumped port (LP). A linear part of Josephson junction (JJ) was simulated by assigning a square patch, connected in parallel to the antenna feeding point, as a parallel LC circuit (LC). LP and LC were connected by two parallel patches. Isometric view of the feeding point of the dipole antenna is depicted in Figure~\ref{fig:fig1}d. The JJ capacitance $c_{J}$ was constant and equal to 10 fF. This value corresponds to the total capacitance of two parallel JJs that form qubit’s SQUID loop with $c_{J}=5$ fF for each junction, which is typical value for the 3D transmon~\cite{RN9}. The qubit resonant frequency was changed by sweeping LC inductance $L_{J}$. All patches, except LC and those connecting it with antenna feeding point, were located on the surface of a substrate with dimensions of 3.5~mm~$\times$~7~mm~$\times$~0.65~mm. The material of all qubit patches was assigned as a perfect conductor while material of the substrate as a sapphire. If the number of qubits was less than the number of cavities in array, blank substrates without qubit were installed into the rest of the cavities.

\section{Inter-cavity coupling}

\begin{figure}
  \centering
  \subfloat{\includegraphics[width=0.49\textwidth]{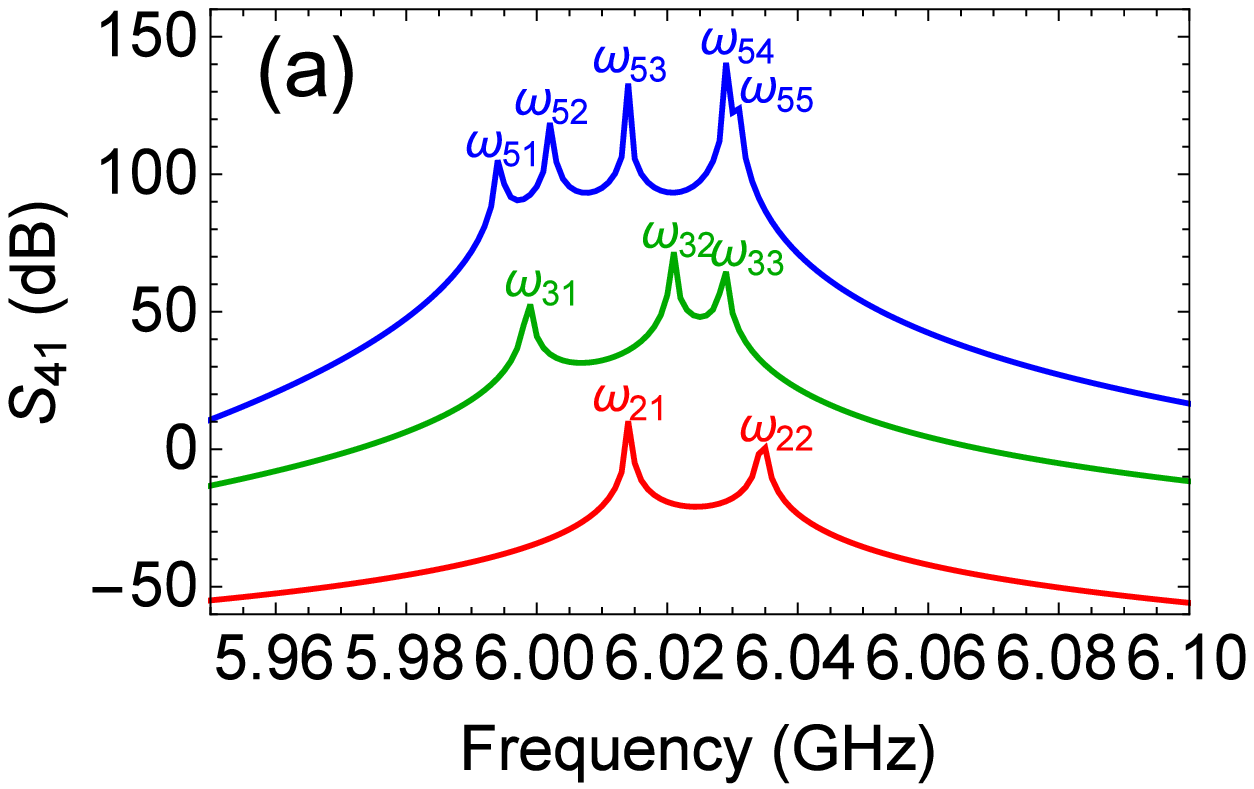}\label{fig:fig2a}}
  \hfill
  \subfloat{\includegraphics[width=0.49\textwidth]{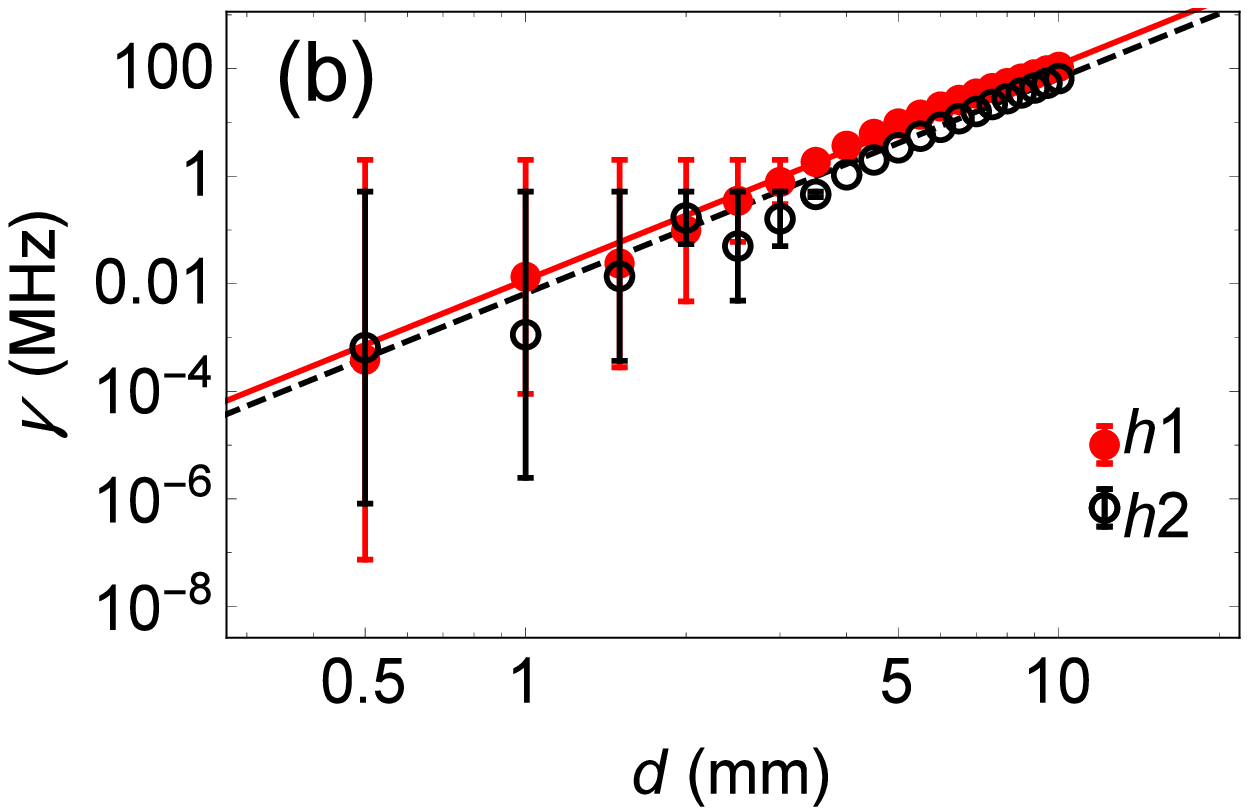}\label{fig:fig2b}}
  \caption{(a) S$_{41}$-parameters for 2- (bottom), 3- (middle) and 5-cavity arrays (top) with $h1$ cavity height. For clarity, middle and top S$_{41}$-parameters are shifted by 50 dB and 140 dB upwards, respectively. (b) Dependence of inter-cavity coupling strength $\gamma$ on coupler diameter $d$ for $h1$ (solid circles) and $h2$ (open circles) cavity heights in 2-cavity array. Solid and dashed lines are fitting functions of $\gamma(d)=\alpha d^{4}$ with $\alpha =11.7$ kHz/mm$^{4}$ and $\alpha=6.6$ kHz/mm$^{4}$ for $h1$ and $h2$, respectively.}
\label{fig:fig2}
\end{figure}

For providing a good coupling between cavities, cavity arrays with different diameters and positions of the coupler were simulated. Figure~\ref{fig:fig2a} shows S$_{41}$-parameters data that were obtained from driven modal simulation of 2-, 3- and 5-cavity arrays with $h1$ cavity height, without qubit and substrate inside. The diameter of the coupler for all arrays was $d=6$ mm with $(0, 0)$ coordinates. As is seen, S$_{41}$-parameters have 2, 3 and 5 resonant peaks for 2-, 3- and 5-cavity arrays, respectively. These peaks correspond to normal mode resonances of cavity array and their frequencies are denoted as $\omega_{ij}$, where $i$ is a number of cavities in array and $j$ is the mode number.

In order to study inter-cavity coupling strength, both eigenmode and driven modal simulations of 2- and 3-cavity arrays were done, from which corresponding normal mode frequencies $\omega_{21}, \omega_{22}, \omega_{31}, \omega_{32}$ and $\omega_{33}$ found. 

Coupling strength between the cavities in 2-cavity array can be characterized by using following frequency model:

\begin{equation}
    \begin{pmatrix}
   \omega_{1} & \gamma\\
    \gamma &\omega_{2}
  \end{pmatrix}\overrightarrow{\nu}=\omega_{\lambda}\overrightarrow{\nu},
  \label{eq:eq1}
\end{equation}
where $\omega_{1, 2}$ are the intrinsic cavity frequencies of cavity 1 and 2, $\gamma$ is the inter-cavity coupling strength and $\omega_{\lambda}=\omega_{21, 22}$ are eigenmode frequencies associated to mode $\overrightarrow{\nu}$. From the spatial symmetry, we can assume that $\omega_{1}=\omega_{2}=\omega$. Thus:

\begin{equation}
  \omega=\omega_{21}+\gamma=\omega_{22}-\gamma, \gamma=(\omega_{22}-\omega_{21})/2.
 \label{eq:eq2}
\end{equation}

For 3-cavity array, the model~\ref{eq:eq1} could be extended to the 3 by 3 matrix:

\begin{equation}
  \begin{pmatrix}
   \omega_{1} & \gamma_{12}&0\\
    \gamma_{12} &\omega_{2}&\gamma_{23}\\
     0 &\gamma_{23}& \omega_{3}
  \end{pmatrix}\overrightarrow{\nu}=\omega_{\lambda}\overrightarrow{\nu},
  \label{eq:eq3}
\end{equation}
where $\omega_{1, 2, 3}$ are intrinsic cavity frequencies of cavity 1, 2 and 3, $\gamma_{12, 23}$ are respectively coupling strengths between cavity 1 and 2 and cavity 2 and 3, while $\omega_{\lambda}= \omega_{31, 32, 33}$. From the spatial symmetry of 3-cavity array, we can assume that $\omega_{1}=\omega_{3}$ and $\gamma_{12}=\gamma_{23}=\gamma$. In this case, eigenmode frequencies could be expressed by using following formulas:

{\small\begin{equation}
\omega_{31}=\frac{1}{2}(\omega_{1}+\omega_{2}-\sqrt{(\omega_{1}-\omega_{2})^{2}+8\gamma^{2}}), 
\omega_{32}=\omega_{1}, 
\omega_{33}=\frac{1}{2}(\omega_{1}+\omega_{2}+\sqrt{(\omega_{1}-\omega_{2})^{2}+8\gamma^{2}}).
 \label{eq:eq4}
\end{equation}}The system of equations~\ref{eq:eq4} has only two unknowns ($\omega_{2}$ and $\gamma$), which can be easily determined.

In the beginning, dependence of inter-cavity coupling strength $\gamma$ on diameter of the coupler $d$ was studied in 2-cavity array. The cavities were empty and the coupler was located at the center of the structure ($(0, 0)$ coordinates). The results of the $\gamma$ calculation by using eigenmode simulation data and~\ref{eq:eq2} for different coupler diameters $d$ are shown in Figure~\ref{fig:fig2b} as solid and open circles for $h1$ and $h2$, respectively. The error bars demonstrate maximum difference between $\gamma$ values for eigenmode and driven modal simulations. As one can see in Figure~\ref{fig:fig2b}, $\gamma$  has a power-law dependence on coupler diameter $d$ and can be approximated by function $\gamma(d)=\alpha d^{4}$ with $\alpha =11.7$ kHz/mm$^{4}$ (solid line) and $\alpha=6.6$ kHz/mm$^{4}$ (dashed line) for $h1$ and $h2$, respectively.

\begin{figure}
  \centering
  \subfloat{\includegraphics[width=0.3\textwidth]{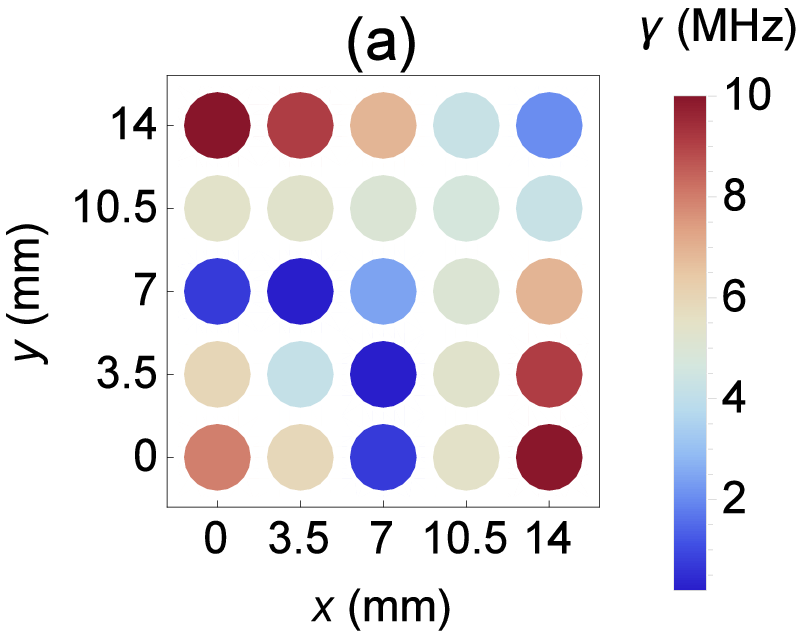}\label{fig:fig3a}}
  \hfill
  \subfloat{\includegraphics[width=0.3\textwidth]{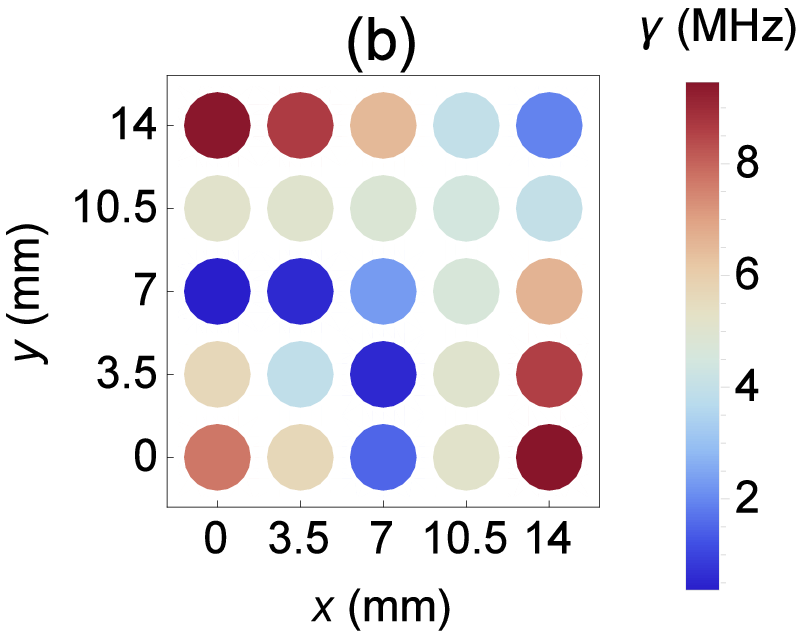}\label{fig:fig3b}}
  \hfill
  \subfloat{\includegraphics[width=0.33\textwidth]{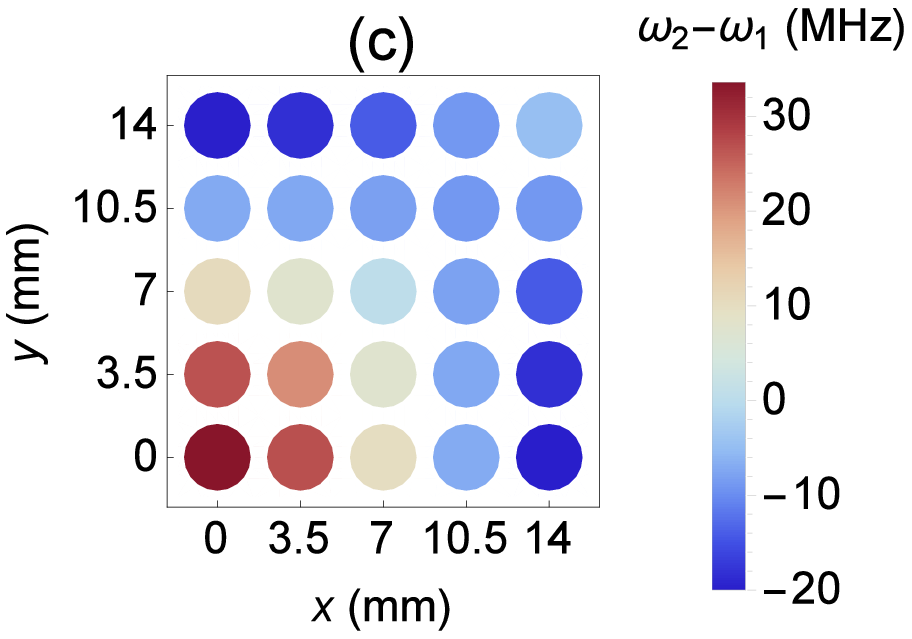}\label{fig:fig3c}}
  \caption{Dependence of inter-cavity coupling strength $\gamma$ on coupler coordinates $(x, y)$ in 2-cavity array~(a) and in 3-cavity array~(b). (c)~The dependence of the difference between intrinsic frequencies $\omega_{2}-\omega_{1}$ on coupler coordinates $(x, y)$ in 3-cavity array. Each disc shows coupler position while its color shows corresponding value of the $\gamma$~(a, b) or $\omega_{2}-\omega_{1}$~(c) according to the color scale bar.  For both arrays, coupler diameter was $d=6$~mm and its position was changed only in the first quadrant of cavity array due to the symmetry of the structures.}
\label{fig:fig3}
\end{figure}

The next step was to determine the position of the coupler where maximum inter-cavity coupling can be achieved. For this purpose, eigenmode simulations of 2- and 3-cavity arrays with fixed coupler diameter $d=6$~mm and different coupler positions were done. The coordinates of the coupler were changed only in the first quadrant of the cavity arrays due to symmetry of the structures. Inter-cavity coupling strengths were calculated by using results of simulations and ~\ref{eq:eq2}, \ref{eq:eq4}. Figures~\ref{fig:fig3a},~\ref{fig:fig3b} show dependences of inter-cavity coupling strength $\gamma$ on coupler position in 2- and 3-cavity arrays. For both cases maximum values of $\gamma$ can be achieved when the coupler is located either at the cavity center ($(0, 0)$ coordinates) or close to the cavity walls ($(0, 14)$ and $(14, 0)$ coordinates). The dependence of the difference between intrinsic frequencies $\omega_{2}-\omega_{1}$ on coupler coordinates in 3-cavity array is shown in Figure~\ref{fig:fig3c}. It is seen that when the coupler is at the center of the cavity, $\omega_{2}>\omega_{1}$, whereas $\omega_{2}<\omega_{1}$ when the coupler is at the edge.

\section{Electrical field distribution in cavity array}\label{sec:sec4}

For getting higher inter-cavity and correspondingly inter-qubit coupling, it is more logical to locate the coupler and qubit at the cavity center $(0, 0)$, where usually TE101 mode of the single rectangular cavity has the maximum of electrical field (E-field). However, in this case the coupler hole destroys uniformity of E-field in the cavity center that might affect qubit-normal mode coupling. Therefore, the better choice would be to locate the coupler either at $(0, 14)$ or $(14, 0)$ coordinates while substrate with the qubit at the cavity center.

\begin{figure}
\includegraphics[width=\linewidth]{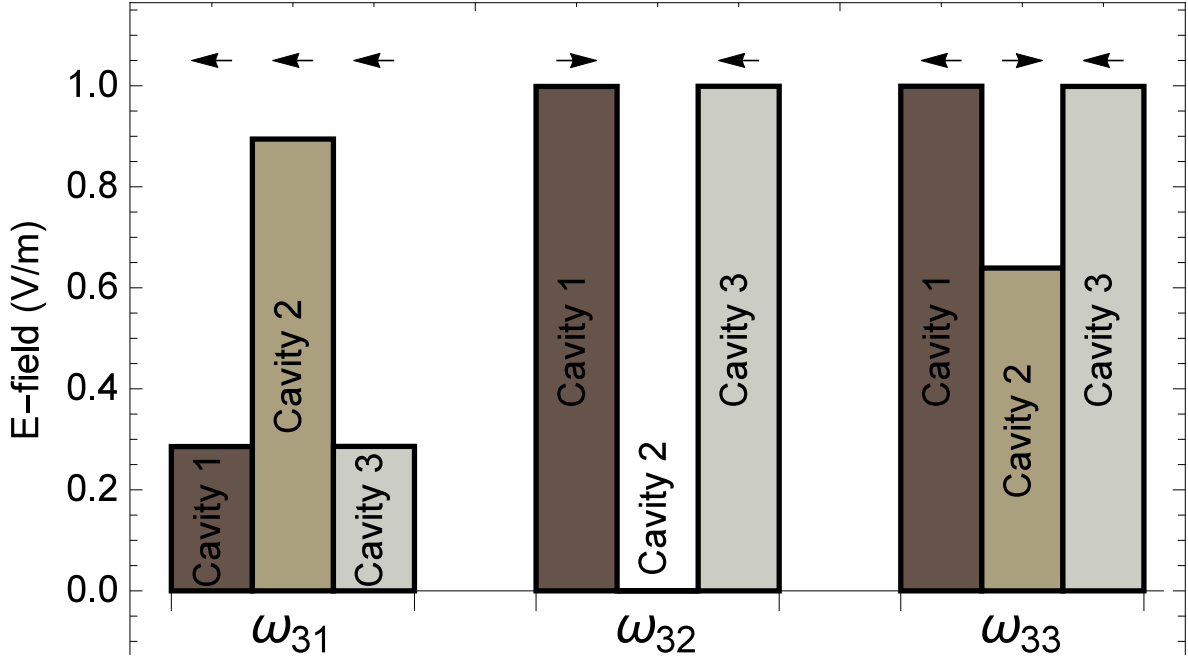}
\caption{
Single photon E-field for 3 normal modes in the cavities of 3-cavity array with coupler coordinates $(0, 14)$. E-field values were taken at the centroid of the cavity cuboids. The data represent only component of E-field which is propagating along the array (the main contributor to the total E-field). Arrows over the bars demonstrate relative directions of the vector of E-field.
}\label{fig:fig4}
\end{figure}

To study E-field, eigenmode simulations of 3-cavity array with coupler coordinates $(0, 14)$ were done. Since the vector of E-field predominantly propagates along the cavity array structure, scalar values of only this component were retrieved and only from the centroid of the cavity cuboids where JJ of the qubit would be located. In HFSS, eigenmode E-field is always normalized to $1$ V/m, for which a stored energy $W=4\times10^{-18}$ Joules in the single cavity was found by using built-in calculator. For cQED experiments, the practical E-field should be rescaled to the single photon level by using the scaling law $E\propto\sqrt{W}$ . The single photon energy for TE101 cavity resonant frequency ($\omega/2\pi=6$ GHz) is $W_{ph}=\hbar\omega=4\times10^{-24}$ Joules, giving the single photon electric field in the order of $10^{-3}$ V/m. A bar chart with rescaled E-field for 3 normal modes in the cavities of 3-cavity array is presented in Figure~\ref{fig:fig4}. Arrows over the bars demonstrate relative directions of the vector of E-field in each cavity of the array for corresponding mode.

It is seen from the bar chart that E-fields for each mode in the cavity 1 and 3 are almost identical, demonstrating a reflection symmetry of the structure. At the same time, E-field in the cavity 2 for mode 2 ($\omega_{32}$) is almost completely suppressed. The reason of such behavior is vectors of E-field in the cavity 1 and 3, which are approximately equal and directed in counter-phase towards each other, suppressing any excitations in the cavity 2, similar to the eigenmodes in coupled pendulums. Suppression of E-field inside different cavities of cavity array can be also found in arrays with number of cavities $N>3$. Thus, $\omega_{32}$ mode cannot excite the transitions of the qubit located in the cavity~2, in other words, the qubit is darkened in respect to this mode. The evidence of the dark state appearance was observed during simulation of qubit in 3-cavity array (see Section~\ref{sec:sec5} and \ref{sec:B}). However, the thorough investigation of the dark state was out of the scope of this paper.

\section{Inter-qubit coupling in qubit-3D cavity array}\label{sec:sec5}

In order to study inter-qubit coupling in qubit-3D cavity array we have considered an array of 3 coupled 3D cavities with two qubits (Q1 and Q2) that were placed inside different cavities of the array. Without loss of generality, we investigated two configurations: nearest-neighbor (NN) and next-nearest-neighbor (NNN). In both configurations, Q1 was in cavity 1 while Q2 was either in cavity 2 for the NN or in cavity 3 for the NNN configuration. For description of these systems, two $5\times5$ Hamiltonian matrices were built based on the Hamiltonian proposed in~\cite{RN13}: 
        
\begin{equation}
  H_{1}=
  \begin{pmatrix}
   \omega & \gamma & 0 & g_{1} & 0\\
   \gamma & \omega & \gamma & 0 & g_{2}\\
   0 & \gamma & \omega & 0 & 0\\
   g_{1} & 0 & 0 & q_{1} & 0\\
   0 & g_{2} & 0 & 0 & q_{2}
  \end{pmatrix},
  \label{eq:eq5}
\end{equation}

\begin{equation}
  H_{2}=
  \begin{pmatrix}
   \omega & \gamma & 0 & g_{1} & 0\\
   \gamma & \omega & \gamma & 0 & 0\\
   0 & \gamma & \omega & 0 & g_{2}\\
   g_{1} & 0 & 0 & q_{1} & 0\\
   0 & 0 & g_{2} & 0 & q_{2}
  \end{pmatrix}.
  \label{eq:eq6}
\end{equation}
\ref{eq:eq5} describes NN and \ref{eq:eq6} NNN configuration. Here $\omega$ is fundamental mode of the single cavity, $q_{1}$($q_{2}$) is Q1(Q2) resonant frequency, $\gamma$ is the inter-cavity and $g_{1}$($g_{2}$) is Q1(Q2)-cavity coupling strengths. It was assumed that $g_{1}=g_{2}=g$ due to the identity of Q1 and Q2. All unknown parameters in~\ref{eq:eq5} and \ref{eq:eq6} were obtained from a series of simple HFSS simulations.  

Single cavity mode $\omega=5.642$ GHz was determined from S$_{41}$-parameter data in driven modal simulation of the single cavity. For this simulation, the substrate with qubit’s antenna and LP but without LC patch, was placed into the cavity. Inter-cavity coupling strength $\gamma=25$ MHz was found by using eigenmode simulation data of the empty 2-cavity array and~\ref{eq:eq2}. The coupler diameter in the simulation was $d=8$ mm with $(0, 13)$ coordinates.

In order to find qubit-cavity coupling $g$, we simulated the single qubit, which was located in the single cavity. This system could be described by using the effective Hamiltonian:

\begin{equation}
  H_{qr}=
  \begin{pmatrix}
   \omega & g\\
   g &q
  \end{pmatrix},
  \label{eq:eq7}
\end{equation}
where $q$ is qubit frequency. Since the qubit is simulated by the LC patch, $q$ can be approximated by the standard formula for LC circuit resonance:

\begin{equation}
  q=\frac{1}{2\pi\sqrt{c_{\Sigma}L_{J}}},
\label{eq:eq8}
\end{equation}
were $c_{\Sigma}$ is the total capacitance of the qubit-cavity system.

\begin{figure}
  \centering
  \subfloat{\includegraphics[width=0.49\textwidth]{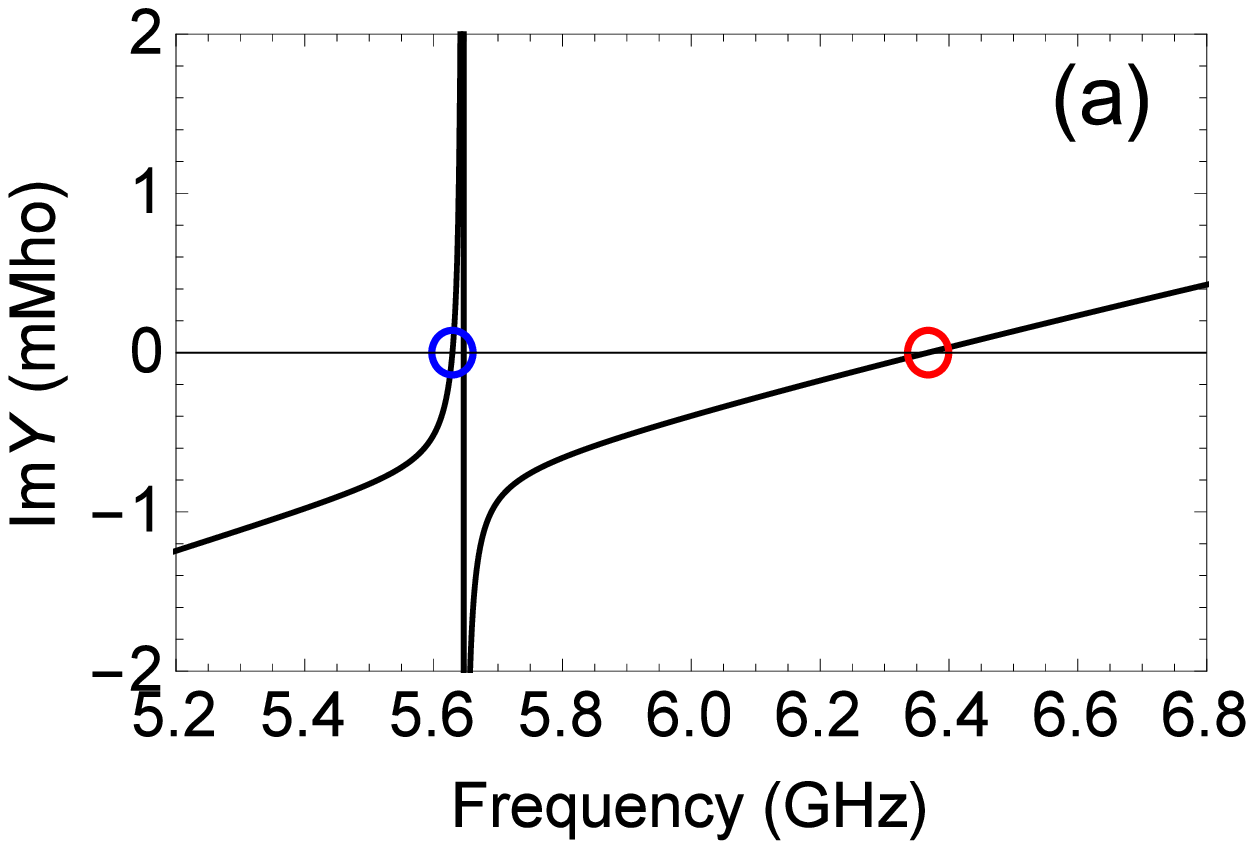}\label{fig:fig5a}}
  \hfill
  \subfloat{\includegraphics[width=0.49\textwidth]{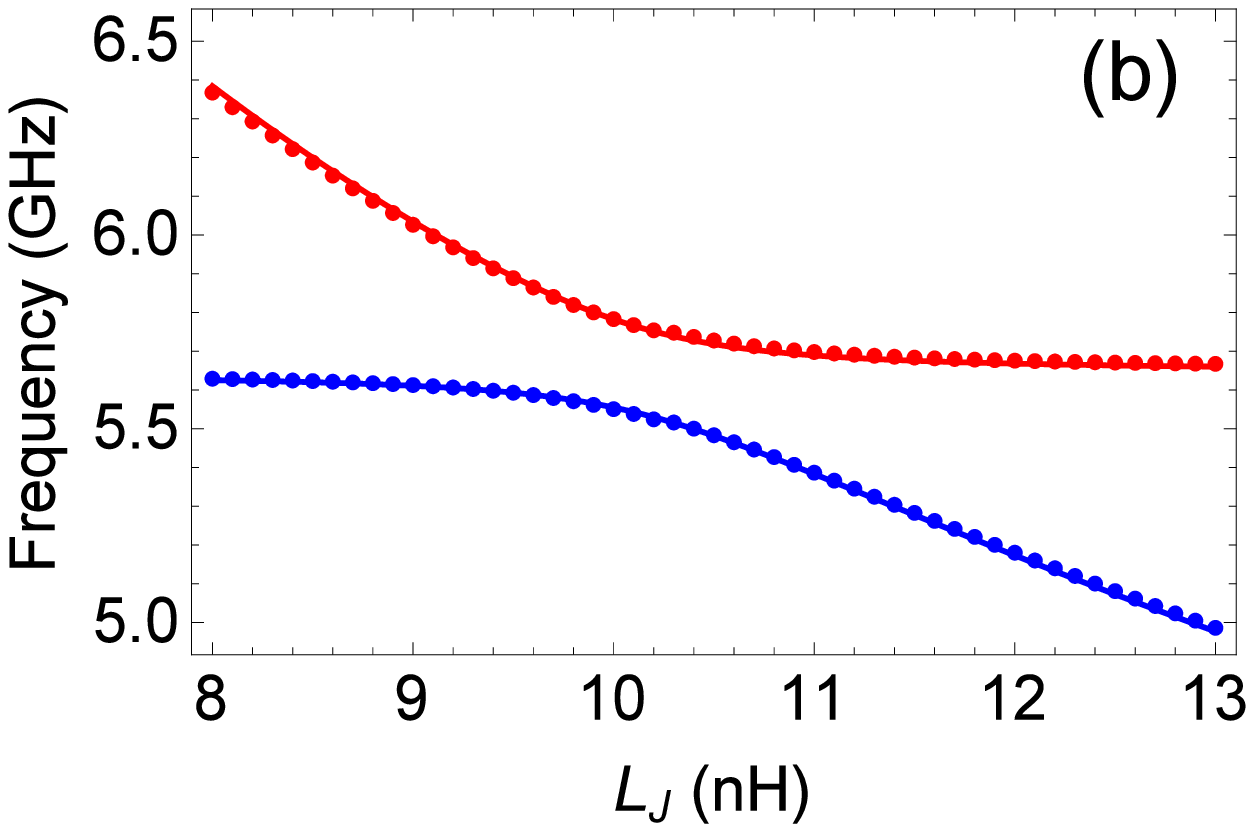}\label{fig:fig5b}}
  \caption{(a) Dependence of imaginary part of admittance Im$Y$ on frequency at the lumped port LP of the qubit, which is located in the single cavity, for $L_{J}=8$~nH. Zeros of Im$Y$ correspond to the cavity (blue open circle) and qubit (red open circle) resonances. (b) An avoided crossing of qubit and cavity resonances, which were obtained from Im$Y$ data by sweeping $L_{J}$. Zeros of Im$Y$ are shown as dots while data fitting with~\ref{eq:eq7} as lines.}
\label{fig:fig5}
\end{figure} 

Figure~\ref{fig:fig5a} shows the imaginary part of admittance Im$Y$ as a function of frequency at the lumped port LP of the qubit for $L_{J}=8$~nH. Zeros of Im$Y$ correspond to resonant frequencies of either qubit or cavity and shown as red and blue open circles, respectively. For $L_{J}=8$~nH, qubit and cavity resonances are far detuned and have a small impact on each other. In this case, we can apply~\ref{eq:eq8} for finding $c_{\Sigma}$. Thus, for $L_{J}=8$~nH, qubit frequency $q=6.368$ GHz, giving $c_{\Sigma}=78$ fF. As a result, we can estimate charging energy $E_{C}=e^{2}/(2c_{\Sigma})=0.248$ GHz and Josephson energy $E_{J}=\phi_{0}^{2}/L_{J}=20.433$ GHz of the qubit, where $\phi_{0}=\hbar/(2e)$ is the reduced flux quantum.

Figure~\ref{fig:fig5b} shows an avoided crossing between qubit and cavity resonant modes, which were obtained from Im$Y$ by sweeping $L_{J}$. Zeros of Im$Y$ are depicted as dots while eigenvalues of~\ref{eq:eq7}, where $g$ was a fitting parameter, are depicted as lines. From the data fitting, the value of the qubit-cavity coupling strength $g=110$ MHz. We should notice that an alternative method, which does not require any fitting parameters and based on HFSS data and BBQ model, gives the same value of $g$. More details about this method can be found in~\ref{sec:A}.

It was found that for such a simple model as the single cavity in the single qubit, it is easy to obtain a good convergence in simulations. Therefore, the same results can be obtained if instead of zeros of Im$Y$, we use poles of imaginary part of inductance Im$Z$. The comparison of Im$Y$ and Im$Z$ data is presented in~\ref{sec:C}.

\begin{figure}
  \centering
  \subfloat{\includegraphics[width=0.305\textwidth]{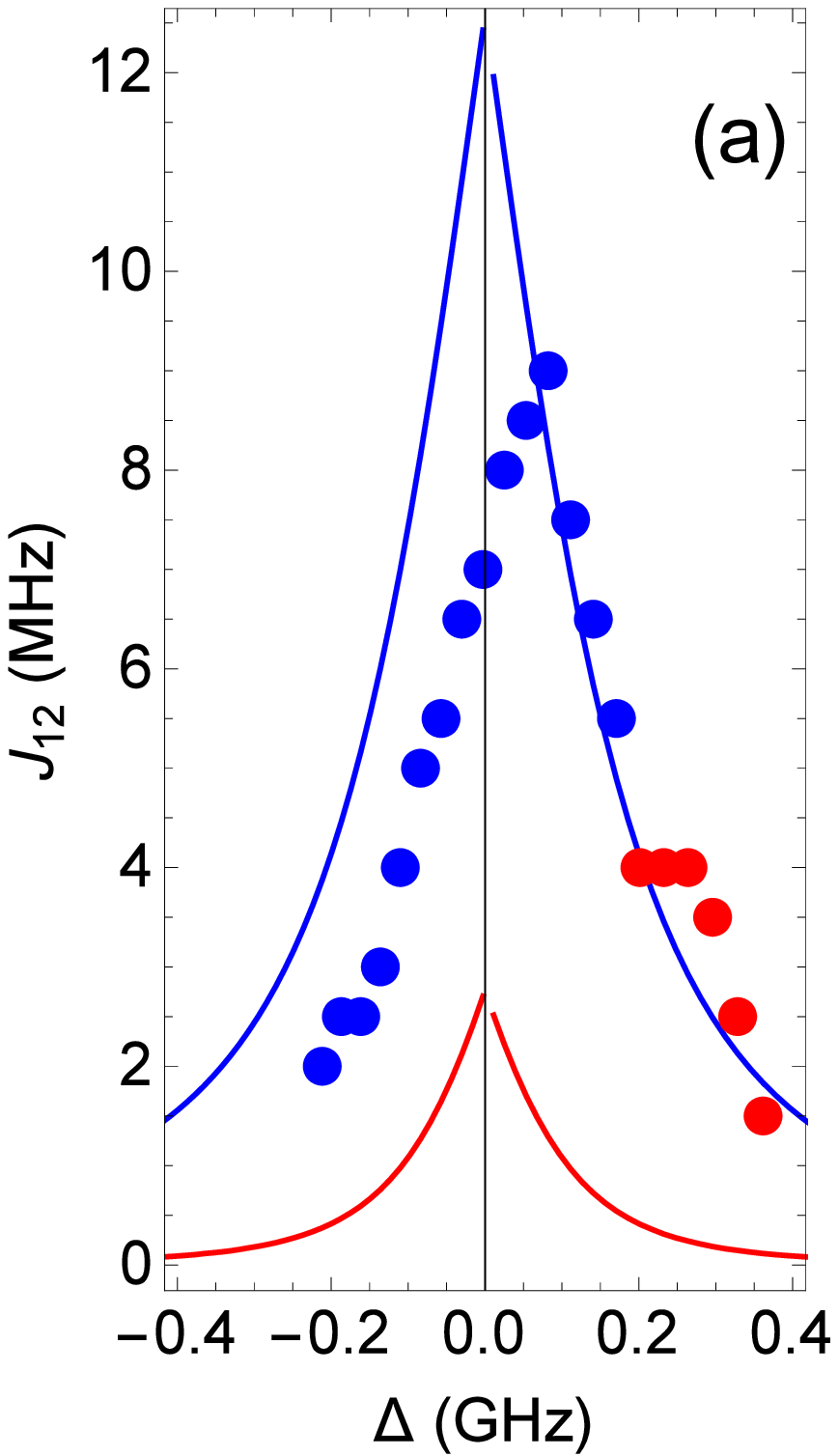}\label{fig:fig6a}}
  \hfill
  \subfloat{\includegraphics[width=0.32\textwidth]{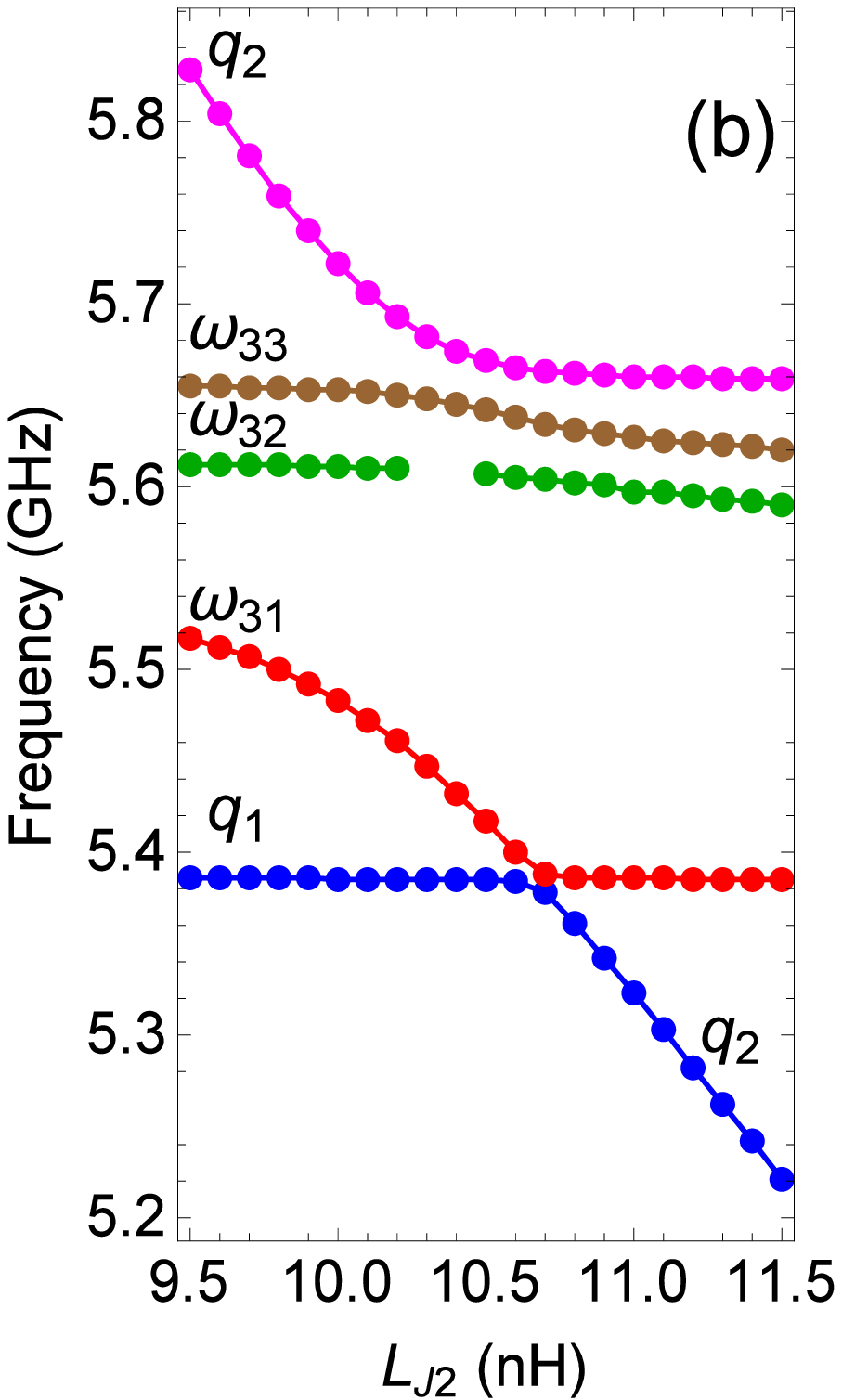}\label{fig:fig6b}}
  \hfill
  \subfloat{\includegraphics[width=0.335\textwidth]{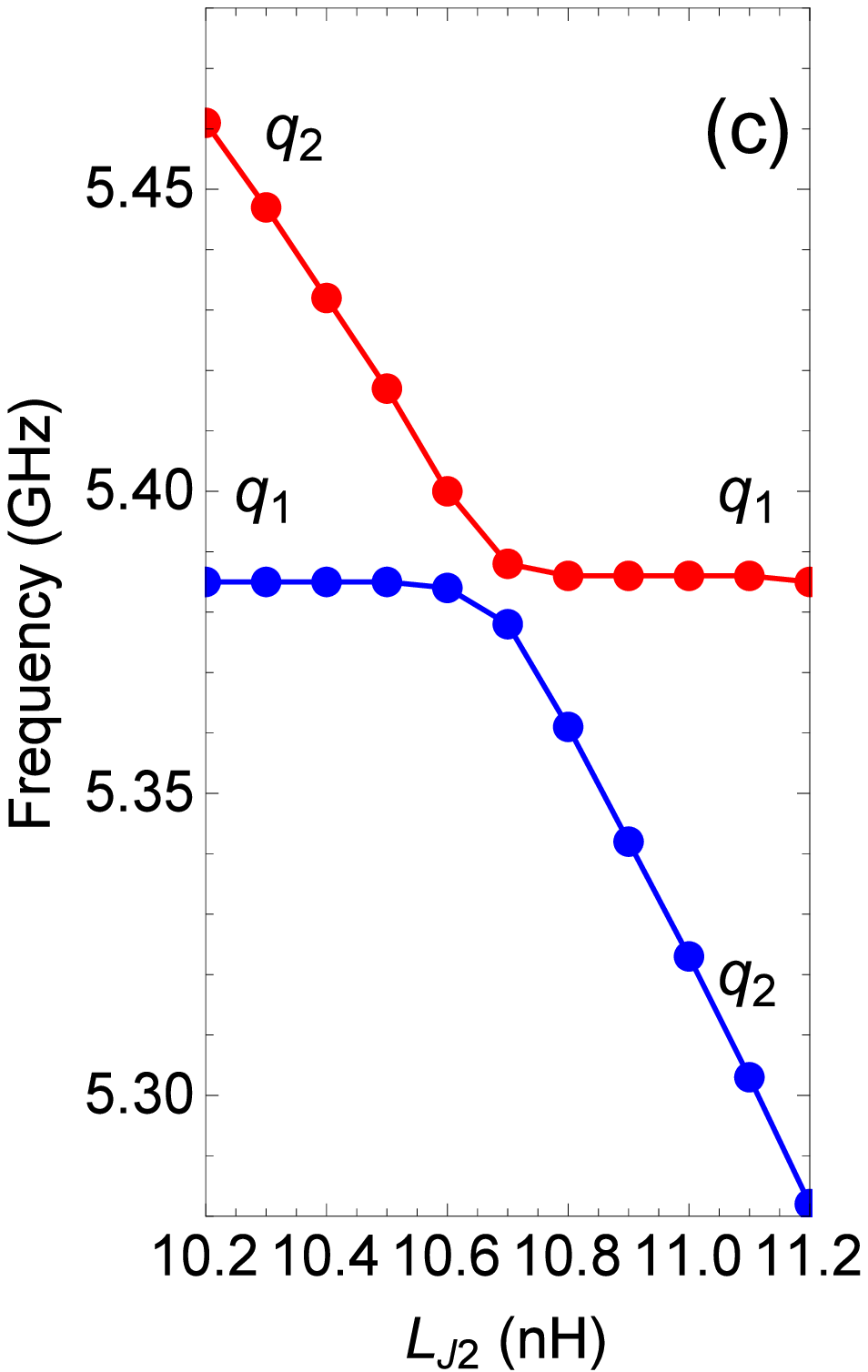}\label{fig:fig6c}}
  \caption{(a) Dependences of the inter-qubit coupling strength $J_{12}$ on detuning $\Delta=q_{1}-\omega$. Results obtained from the model Hamiltonians~\ref{eq:eq5} and \ref{eq:eq6} are shown as blue and red lines while those obtained from HFSS simulations as blue and red circles for the NN and NNN configurations, respectively. (b) Distribution of energy spectrum for the NN configuration. The data were extracted from Im$Z$ at the LP of Q1 for fixed inductance $L_{J1}=10.5$~nH of Q1 by sweeping inductance $L_{J2}$ of Q2. (c) Magnified region of avoided crossing between $q_{1}$ and $q_{2}$, which is shown in (b).}
\label{fig:fig3}
\end{figure}

After finding all the unknown parameters, \ref{eq:eq5} and \ref{eq:eq6} were diagonalized. As a result, inter-qubit coupling strengths $J_{12}$ were determined as the half of the minimum distance between $q_{1}$ and $q_{2}$ modes. The dependences of $J_{12}$ on detuning of qubit from the single cavity mode $\Delta=q_{1}-\omega$ are shown in Figure~\ref{fig:fig6a} as blue and red lines for the NN and NNN configurations, respectively. 

The validity of the model Hamiltonian was verified by using driven modal simulations of 3-cavity array with two identical qubits (Q1 and Q2) in the same configurations. The coupler diameter was $d=8$ mm with $(0, 13)$ coordinates. For different detunings $\Delta$ of Q1, the frequency $q_{2}$ of Q2 was swept in the vicinity of $q_{1}$ and modes of the cavity array. Unfortunately, achieving a good convergence even for such not very complex models requires long computation times, which we tried to avoid by reducing our demands to the convergence. Therefore, $J_{12}$ data in our simulation was possible to extract not for every $\Delta$ and only from Im$Z$.  Poles of Im$Z$ were extracted at the LP of Q1. Figure~\ref{fig:fig6b} shows energy spectrum for the NN configuration, which was obtained from Im$Z$ data by sweeping inductance $L_{J2}$ of Q2 at $L_{J1}=10.5$~nH. The absence of few points for $\omega_{32}$ in Figure~\ref{fig:fig6b} is related to the dark state of Q2 at this mode (see Section~\ref{sec:sec4} and \ref{sec:B}). As is seen, $q_{1}$ and $q_{2}$ frequencies have an avoided crossing, magnified region of which is depicted in Figure~\ref{fig:fig6c}. Inter-qubit coupling strength $J_{12}$ was estimated as the half of the minimum frequency difference between $q_{1}$ and $q_{2}$ data points. The dependence of $J_{12}$ on $\Delta$ is shown in Figure~\ref{fig:fig6a} as blue and red circles for the NN and NNN configurations, respectively. As it well seen from the NN configuration, a zero detuning ($\Delta=0$) symmetry of the $J_{12}$ data for the simulation and model Hamiltonian is shifted. The cause of the shift are different total capacitances $c_{\Sigma}$ for two approaches. For the Hamiltonian diagonalization, we used $c_{\Sigma}$ of the single cavity and single qubit system while for the simulation, $c_{\Sigma}$ in the 3-cavity array and 2 qubits system is different. This difference leads to the different calculated and simulated $q_{1}$ frequencies for the same $L_{J1}$, giving different detunings $\Delta$. Despite this fact, the $J_{12}$ values for the model and simulation demonstrate a good agreement. For the NN configuration, the maximum values of the inter-qubit coupling strength $J_{12}=12.4$~MHz and $J_{12}=9$~MHz while for the NNN configuration $J_{12}= 2.7$~MHz and $J_{12}=4$~MHz for the model and simulation, respectively. As one can see, the difference between two methods for both configurations is approximately $30\%$. When $\Delta\gg g, \gamma$, analytical results for NN and NNN configurations are respectively $J_{12}=2g^{2}\gamma/\Delta^{2}$ and $J_{12}=2g^{2}\gamma^{2}/\Delta^{3}$, so the inter-qubit coupling strength for NN configuration is generally bigger than for NNN.

\section{Conclusions}

We have presented the practical design of linear array of three-dimensional (3D) cavities for experiments in circuit quantum electrodynamics. In order to obtain an efficient inter-cavity, qubit-cavity and inter-qubit coupling, geometry of the structure was optimized by using high frequency structure simulator (HFSS). A method based on Hamiltonian description of the qubit-3D cavity array system for prediction of inter-qubit interaction mediated by normal modes of 3D cavity array was proposed. Unknown parameters in the model Hamiltonian could be found from a series of simple HFSS simulations. The validity of the method was confirmed from direct observation of inter-qubit coupling in simulations, which demonstrate a good agreement. The results obtained allow us to propose this technique for determination of inter-qubit coupling in more complex qubit-cavity systems where finite-element electromagnetic simulators require huge computational resources.

\ack

The authors are grateful to the National Center for High-Performance Computing for computer time and facilities. Also, the authors thank Chii-Dong Chen, Cen-Shawn Wu, Yu-Han Chang and Raveendharan Sundaram for valuable discussions. This work is financially supported by the Ministry of Science and Technology, Taiwan under grant No107-2112-M-005-001. 

\appendix\

\section{Calculation of qubit-cavity coupling strength by using BBQ model}\setcounter{section}{1}\label{sec:A}

BBQ provides an alternative approach for calculation of qubit-cavity coupling strength $g$, which is based on HFSS simulation data \cite{RN9}. 
The only information needed is the dependence of admittance Im$Y$ on frequency at LP for the case when qubit is positively far detuned from the cavity resonance. According to the BBQ model, self-Kerr susceptibility $\chi_{p}$ of qubit or cavity mode is:

\begin{equation}
 \chi_{p}=-\frac{L_{p}e^{2}}{2L_{J}c_{p}},
\label{eq:eqB1}
\end{equation}
where $L_{p}$ and $c_{p}$ are respectively inductance and capacitance of qubit or cavity mode $\omega_{p}$. $L_{p}$ and $c_{p}$ could be found from the following relations:

\begin{equation}
L_{p}=\frac{1}{\omega_{p}^{2}c_{p}},
\label{eq:eqB2}
\end{equation}

\begin{equation}
c_{p}=\frac{1}{2}\rm Im\it \frac{dY(\omega_{p})}{d\omega_{p}}.
\label{eq:eqB3}
\end{equation}
Using simulated data of Im$Y$ for $L_{J}=8$~nH (see main text), self-Kerr susceptibilities $\chi_{q}=–227$ MHz and $\chi_{r}=–0.134$ MHz for the qubit and cavity mode, respectively, could be estimated. Cross-Kerr susceptibility, which is $\chi_{qr}=-2\sqrt{\chi_{q}\chi_{r}}$ is related to the qubit-cavity coupling strength $g$ as~\cite{RN14}:

\begin{equation}
\chi_{qr}=-E_{C}\frac{g^{2}}{\Delta_{qr}^{2}},
\label{eq:eqB4}
\end{equation}
where $\Delta_{qr}$ is detuning between qubit and cavity mode. Taking into account that for $L_{J}=8$~nH, $\Delta_{qr}=739$~MHz, we get $g=110$~MHz, which is the same as those obtained from the data fitting in the main text.

\section{Dark state}\setcounter{section}{2}\label{sec:B}

\begin{figure}
  \centering
  \subfloat{\includegraphics[width=0.49\textwidth]{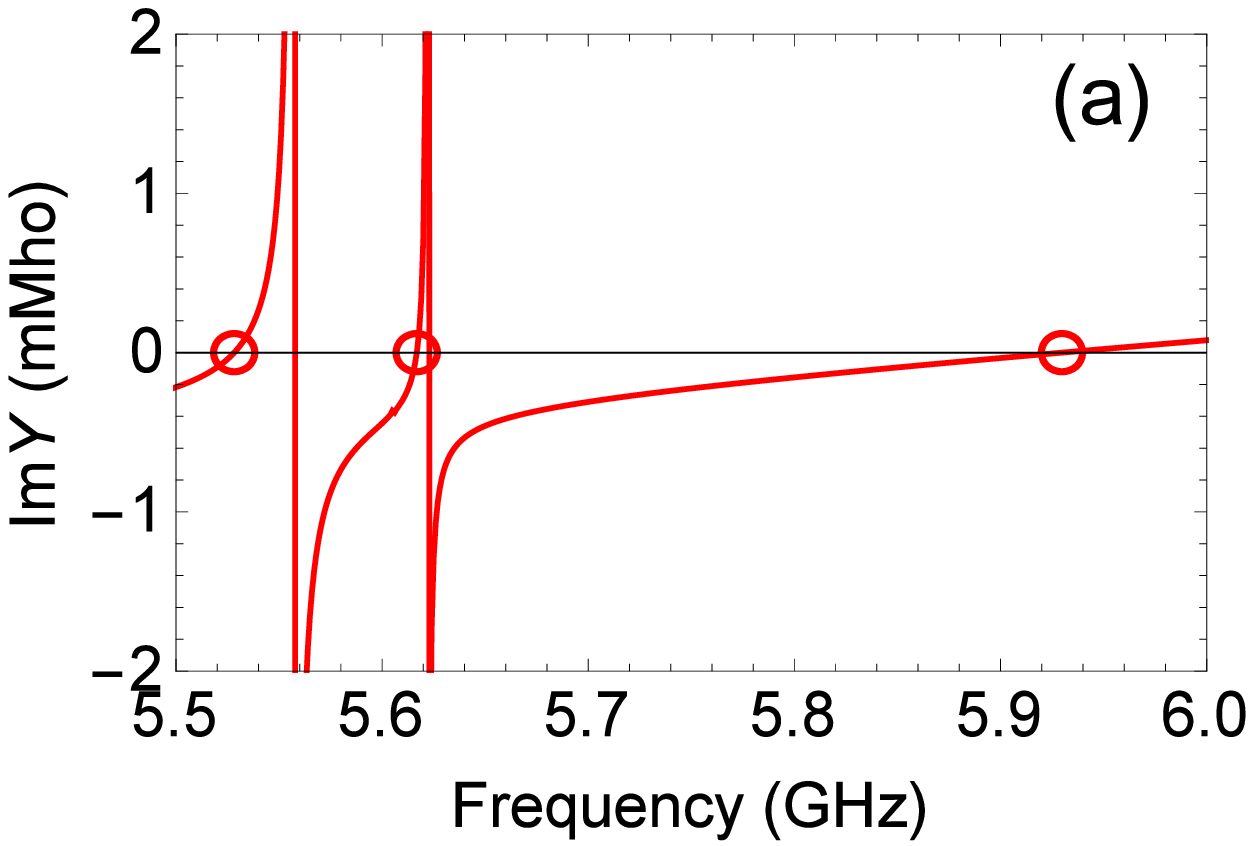}\label{fig:figB1a}}
  \hfill
  \subfloat{\includegraphics[width=0.49\textwidth]{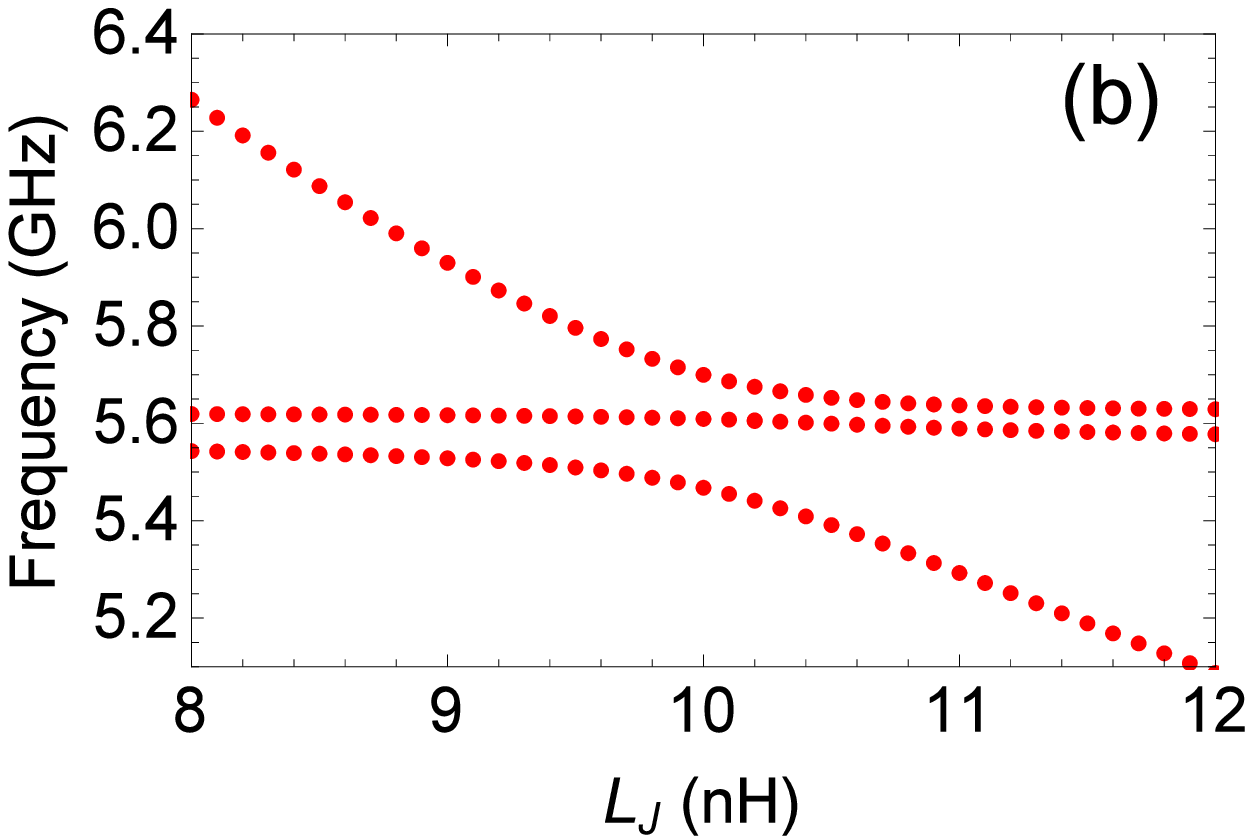}\label{fig:figB1b}}
  \hfill
  \subfloat{\includegraphics[width=0.49\textwidth]{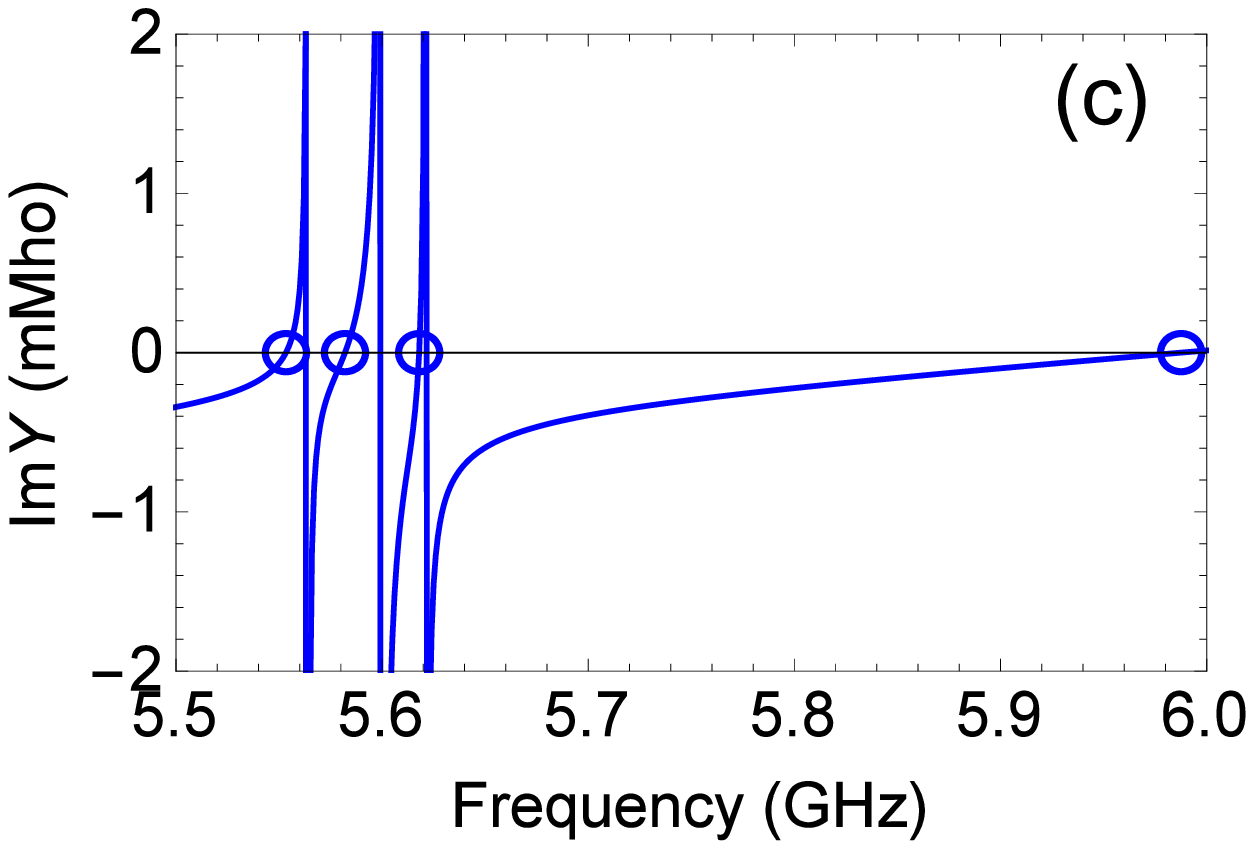}\label{fig:figB1c}}
  \hfill
  \subfloat{\includegraphics[width=0.49\textwidth]{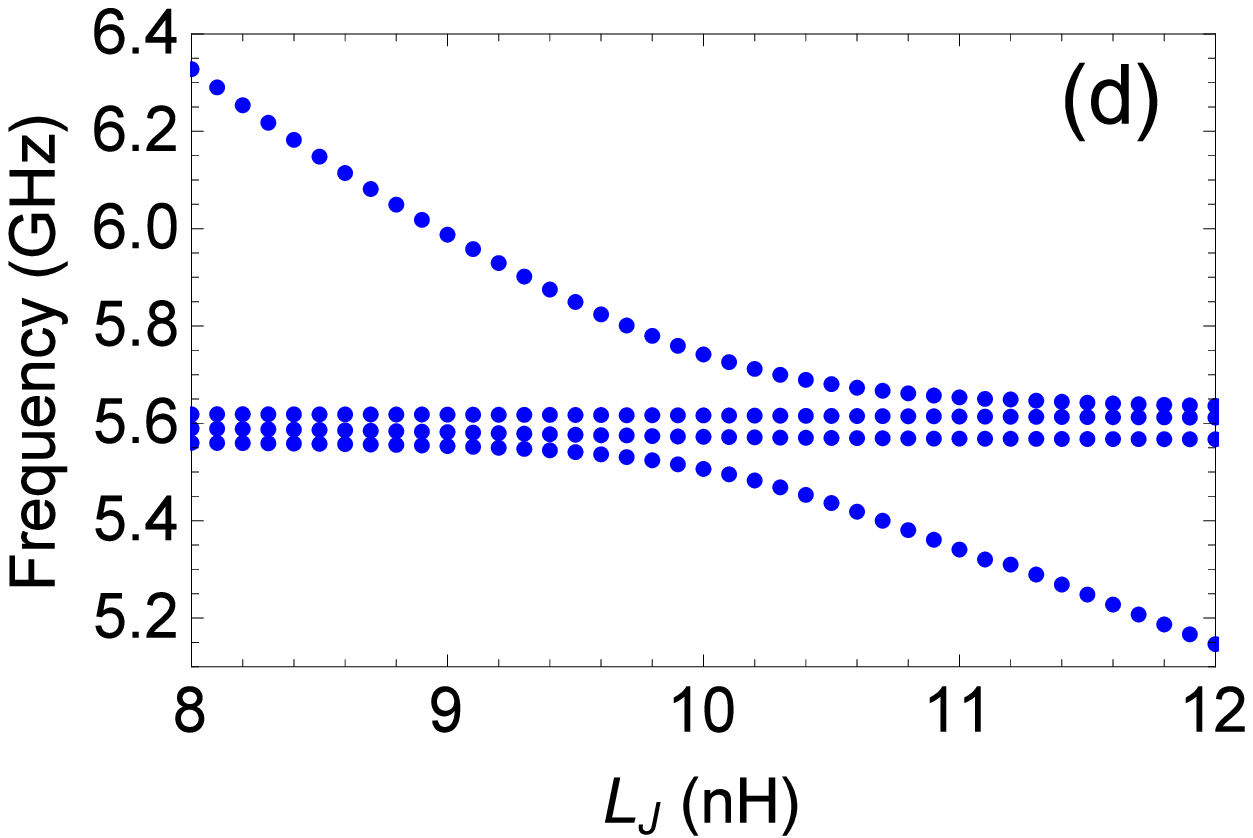}\label{fig:figB1d}}
  \caption{(a) Dependence of imaginary part of admittance Im$Y$ on frequency at the lumped port LP of the qubit located in cavity 2 of 3-cavity array for $L_{J}=9$~nH. (b) Avoided crossings of qubit frequency and 2 normal modes ($\omega_{31}$ and $\omega_{33}$) of 3-cavity array. (c) Dependence of imaginary part of admittance Im$Y$ on frequency at the lumped port LP of the qubit, located in cavity 1 of 3-cavity array for $L_{J}=9$~nH. (d) Avoided crossings of qubit frequency and 3 normal modes ($\omega_{31}$, $\omega_{32}$ and $\omega_{33}$) of 3-cavity array.}
\label{fig:figB1}
\end{figure}

The absence of E-field in the cavity 2 for the 2nd normal mode $\omega_{32}$ in 3-cavity array (see Figure~\ref{fig:fig4} in the main text) must lead to the absence of the coupling between qubit located in cavity 2 and $\omega_{32}$ mode. This effect could be easily revealed by applying the driven modal simulations to 3-cavity array with the qubit located in the cavity 2. The coupler diameter was $d=8$ mm with $(0, 13)$ coordinates. Figure~\ref{fig:figB1}(a) shows dependence of imaginary part of admittance Im$Y$ on frequency at the lumped port LP of the qubit for $L_{J}=9$~nH, with zeros depicted as open circles. Zero associated with the qubit mode is located at 5.93 GHz while two zeros related to two normal modes ($\omega_{31}$ and $\omega_{33}$) are observed at 5.528 GHz and 5.617 GHz. Zeros of Im$Y$ for different values of qubit inductance $L_{J}$ present an avoided crossing, which is shown in Figure~\ref{fig:figB1b}. The absence of the qubit coupling with $\omega_{32}$ mode almost in the whole range of qubit frequencies was observed. For comparison, in Figures~\ref{fig:figB1c} and \ref{fig:figB1d}, simulation data for the qubit located in the cavity 1 present 3 zeros, which are related to 3 normal modes ($\omega_{31}$, $\omega_{32}$ and $\omega_{33}$) as well as zero related to the qubit mode.

\section{Comparison of Im$Y$ and Im$Z$ data}\setcounter{section}{3}\label{sec:C}

\begin{figure}
  \includegraphics[width=\linewidth]{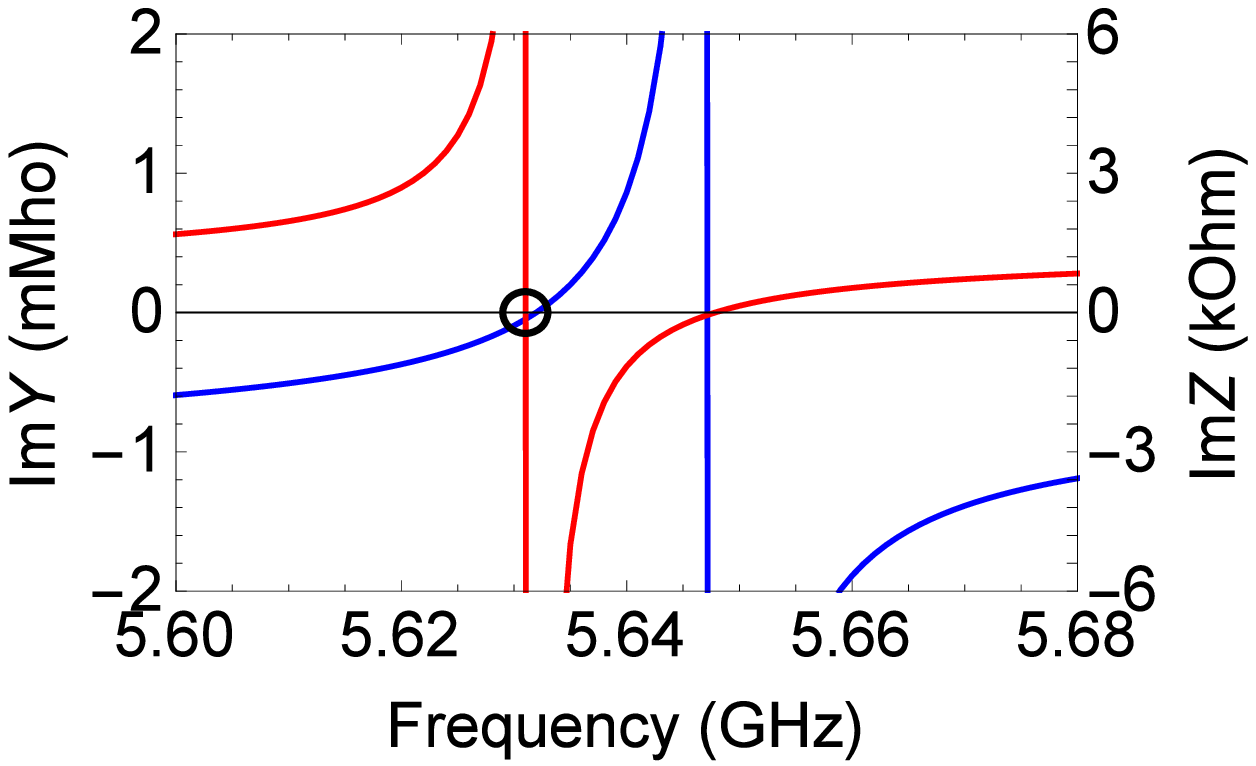}
  \caption{Dependence of imaginary part of admittance Im$Y$ (blue line and left y-axis) and impedance Im$Z$ (red line and right y-axis) on frequency at the lumped port LP of the qubit, which is located in the single cavity, for $L_{J}=8$~nH. Zero of Im$Y$ and pole of Im$Z$, both indicating the cavity resonance, is depicted as a black open circle.}
\label{fig:figC1}
\end{figure}

In most of our simulations, imaginary part of admittance Im$Y$ and impedance Im$Z$ data demonstrate a good convergence and interchangeability. For example, Figure~\ref{fig:figC1} shows dependence of Im$Y$ and Im$Z$ on frequency at the lumped port LP of the qubit for $L_{J}=8$~nH. The simulation was performed for the the same model, results of which are depicted in Figure~\ref{fig:fig5}. As one can see, zero of Im$Y$ and pole of Im$Z$, which indicate the cavity resonance (black open circle in Figure~\ref{fig:figC1}), almost coincide. The difference between Im$Y$ and Im$Z$ data for the qubit and cavity resonances was less than $1$~MHz, which is a typical step for frequency sweep in our simulations.

\section*{References}
\bibliography{ref_text_1} 

\end{document}